\documentclass[a4paper, nofootinbib, showpacs, amsmath, amssymb]{revtex4}

\usepackage{graphicx}

\begin{document}

\title{Minimum-length deformed QM/QFT, issues and problems }

\author{Michael~Maziashvili}
\email{maziashvili@gmail.com} \affiliation{Particle Physics $\mathbf{\&}$ Cosmology Group, Ilia State University, 3/5 Cholokashvili Ave., Tbilisi 0162, Georgia}

\author{and Luka~Megrelidze}

\email{luka.megrelidze.1@iliauni.edu.ge} \affiliation{Particle Physics $\mathbf{\&}$ Cosmology Group, Ilia State University, 3/5 Cholokashvili Ave., Tbilisi 0162, Georgia}

\begin{abstract}

Using a particular Hilbert space representation of minimum-length deformed quantum mechanics, we show that the resolution of the wave-function singularities for strongly attractive potentials, as well as cosmological singularity in the framework of a minisuperspace approximation, is uniquely tied to the fact that this sort of quantum mechanics implies the reduced Hilbert space of state-vectors consisting of the functions nonlocalizable beneath the Planck length. (Corrections to the Hamiltonian do not provide such an universal mechanism for avoiding singularities.) Following this discussion, as a next step we take a critical view of the meaning of wave-function in such a quantum theory. For this reason we focus on the construction of current vector and the subsequent continuity equation. Some issues gained in the framework of this discussion are then considered in the context of field theory. Finally, we discuss the classical limit of the minimum-length deformed quantum mechanics and its dramatic consequences.

\end{abstract}


%04.60.Bc Phenomenology of quantum gravity

\maketitle

\section{Introduction}
\label{intro}

The idea of fundamental length in the context of quantum theory is almost as old as the quantum mechanics itself \cite{Blokhintsev:1973su}. In what follows we will focus on the approach based on the modified position-momentum uncertainty relation. To our knowledge, this sort of approach was originally proposed by Saavedra and Utreras \cite{Saavedra:1979gc, SaavedraUtreras, Saavedra:1980ss, Giffon:1983bg, Talukdar:1982qc,  Montecinos:1985zs} purely in the context of a high energy physics, however, it became popular after its "derivation" in the context of string theory \cite{Veneziano:1986zf, Gross:1987ar, Amati:1988tn, Veneziano:1989fc, Konishi:1989wk, Guida:1990st, Veneziano:1989ti, Witten:1996, Duff:2001ba}. Later on it was replaced by another uncertainty relation universally valid for strings as well as D-branes \cite{Yoneya:2007zza}. So, presently the modified position-momentum uncertainty relation is motivated mainly by the black hole physics in view of combining the basic principles of quantum theory and general relativity \cite{Mead:1964zz, Mead:1966zz, Garay:1994en, Maggiore:1993rv, Scardigli:1999jh, Adler:1999bu}. The construction of a particular Hilbert space representation resulting in the Planck length modified position-momentum uncertainty relation \cite{Kempf:1994su} has sparked considerable interest among the physicists as a reasonable model for studying the quantum gravity phenomenology \cite{Tkachuk:2013qa, Pedram:2012my, Dey:2012tv, Nozari:2012nf, Dey:2012dm, Nozari:2012gd, Sprenger:2012uc, Sprenger:2011jc, Maslowski:2012aj, Mimasu:2012np, Bleicher:2011uj, Chang:2011jj, Sprenger:2010dg, Mimasu:2011sa, Nicolini:2011nz, Kober:2010um, Pedram:2011xj,  Bleicher:2010qr, Quesne:2009vc, Lubo:2004de}.

One of the useful points for discussing the minimum-length deformed quantum mechanics (ml-QM) is the possible avoidance of various singularities in physics like: the singular behaviour of propagators on the light-cone that causes UV divergences in QFT; black-hole and cosmological space-time singularities; the quantum-mechanical singularities of the wave-function in strongly attractive potentials. The idea is based largely on intuition from uncertainty principle and fundamental length scale arguments. Addressing some of these questions from different points of view, at the same time we examine the technical subtleties residing in the mathematical structure of ml-QM. In our discussion we will follow a particular representation of ml-QM constructed in \cite{Kempf:1996nk}, which will be described in the next section. It is important to notice that we do not have even a heuristic argument about the equivalence of different representations for ml-QM (something like of Stone-von Neumann theorem in standard quantum mechanics \cite{SvN}). Going further, we address some of the conceptual points related to the interpretation of the wave function as well as the coupling of the charged field to the electromagnetic one and the question of gauge invariance. Last section is devoted to the classical limit of ml-QM. While in the deep UV region the radical changes are naturally expected in the framework of such a theory, in IR regime one usually assumes the corrections should be strongly suppressed. But unfortunately this is not the case, classical physics seems to be not oblivious to this sort of modification. Throughout the paper, we will use the abbreviation ml-QFT for QFT which follows from the ml-QM.

\section{The structure of ml-QM}

The ml-QM emanates from the deformed position-momentum uncertainty relation 

\begin{equation}\label{grur} \delta X \delta P  \,\geq \,
  \frac{1}{2} \,+\, \beta \delta P^2~,  \end{equation} where $\sqrt{\beta}$ is of the order of the Planck length $l_P \approx 10^{-33}$cm \cite{Veneziano:1986zf, Maggiore:1993rv, Adler:1999bu, Scardigli:1999jh}. The relation \eqref{grur} reflects the onset of gravitational effects in $\mathsf{QM}$ when the energy approaches the quantum gravity scale $\delta P \gtrsim l_P^{-1}$. An important feature of Eq.\eqref{grur} is that it exhibits a lower bound for position uncertainty. The construction of the Hilbert space representation for this sort of quantum mechanics \cite{Kempf:1994su, Kempf:1996nk} has stimulated a great deal of research. One can find the following representation for such a quantum mechanics 
  
  \begin{equation} [\widehat{X},\,\widehat{P}] = i (1 \,+\, \beta \widehat{P}^2)~, \nonumber  \end{equation} which can be solved for $\widehat{X},\,\widehat{P}$ in terms of the standard $\widehat{x},\,\widehat{p}$ operators \cite{Kempf:1996nk} 

\begin{eqnarray}\label{defxopo} && \widehat{X} = \widehat{x}\,,~~ \widehat{P} =  \beta^{-1/2}\tan\left(\widehat{p}\sqrt{\beta}\right) ~.\end{eqnarray} This equation readily indicates the presence of cut-off: $p < \pi /2\sqrt{\beta}$.

\noindent The multidimensional generalization of Eq.\eqref{grur} maybe written as \cite{Kempf:1996nk}

\begin{eqnarray} \left[\widehat{X}^i,\,\widehat{P}^j\right] = i\left( \frac{2\beta \widehat{\mathbf{P}}^2}{\sqrt{1+4\beta \widehat{\mathbf{P}}^2} \,-\, 1}\,\delta^{ij} +2\beta \widehat{P}^i\widehat{P}^j  \right)~, ~~ \left[\widehat{X}^i,\,\widehat{X}^j\right] = \left[\widehat{P}^i,\,\widehat{P}^j\right]=0~.\label{minlengthqm}  \end{eqnarray} The deformed $\widehat{\mathbf{X}},\,\widehat{\mathbf{P}}$ operators in Eq.\eqref{minlengthqm} can be represented in terms of the standard $\widehat{\mathbf{x}},\,\widehat{\mathbf{p}}$ operators in the following way 

\begin{equation}\label{deformedopmultid} \widehat{X}^i = \widehat{x}\,^i\,,~~~~\widehat{P}\,^i = \frac{\widehat{p}\,^i}{1-\beta \widehat{\mathbf{p}}^2}~.  \end{equation} Its Hilbert space realization in the standard-momentum, $\mathbf{p}$, representation has the form

\begin{equation}\label{standardprepresentation} \widehat{X}^i\psi(\mathbf{p}) = i\partial_{p_i} \psi(\mathbf{p})\,,~~~~ \widehat{P}^i\psi(\mathbf{p}) = \frac{p^i}{1-\beta \mathbf{p}^2}\,\psi(\mathbf{p})~, \nonumber \end{equation}  with the scalar product 

\begin{equation}\label{scalarproduct} \langle \psi_1 | \psi_2 \rangle = \int\limits_{\mathbf{p}^2 < \beta^{-1}}d^3p \,\psi^*_1(\mathbf{p})\psi_2(\mathbf{p})~. \nonumber \end{equation} Let us notice that the cutoff $\mathbf{p}^2 < \beta^{-1}$ results from the fact that the whole space $\mathbf{P} \in \mathbb{R}^3$ is covered by the new coordinates $\mathbf{p}$ within the ball $\mathbf{p} \in \mathbb{B}_{\beta} = \left\{\mathbf{p}:\, p < \beta^{-1/2} \right\}$, see Eq.\eqref{deformedopmultid}. Or in other words, when $p$ runs over the region $\left[0, \beta^{-1/2}\right)$, $P$ covers the whole region $[0, \infty)$. It is worth noticing that the existence of this cutoff has an implicit reference to modified Heisenberg's uncertainty relation Eq.\eqref{grur}. Namely, the standard uncertainty relation can be understood on the basis of Fourier transform since the spatial and momentum wave functions are related through it. The Fourier transform has the property that the more tightly localized the spatial wave function is, the less tightly localized the momentum function must be; and vice versa. Consequently, as in the modified theory the momentum wave function can not be wider than $\sim \beta^{-1/2}$, the spatial wave function can not be localized beneath the region $\sim \beta^{1/2}$.

To be more concrete, one can prove the following statement. If the normalized function $\phi(\mathbf{x})$ (that is, $\int d^3x \, \phi^*(\mathbf{x})\phi(\mathbf{x}) =1$) admits the representation  \begin{equation}\label{chamochrilitsarmodgena} \phi(\mathbf{x}) \,=\, \frac{1}{\left(2\pi\right)^{3/2}} \int\limits_{\mathbf{p}^2 < \beta^{-1}} d^3p \,\, e^{-i\mathbf{p}\cdot\mathbf{x}} \psi(\mathbf{p})~, \end{equation} then 

\begin{eqnarray}
\int d^3x \, \phi^*(\mathbf{x}) \left( x^i \,-\, \left\langle x^i \right\rangle \right)^2\phi(\mathbf{x}) \, \geq \, \frac{15 \beta}{16\pi}~, \nonumber \end{eqnarray} where, as usual, $\langle x\rangle \equiv \int d^3x \, \phi^*(\mathbf{x}) x^i \phi(\mathbf{x})$. Namely, with no loss of generality, one can assume  \[ \left\langle p^i \right\rangle \,=\, \int\limits_{\mathbf{p}^2 < \beta^{-1}} d^3p \,\, \psi^*(\mathbf{p})p^i\psi(\mathbf{p}) \,=\, 0  ~, \] then the standard uncertainty relation for the Fourier transform pair takes the form

\begin{eqnarray}
\int d^3x \, \left( x^i \,-\, \left\langle x^i \right\rangle \right)^2 \left|\phi(\mathbf{x})\right|^2 \, \geq \, \frac{1}{4 \int\limits_{\mathbf{p}^2 < \beta^{-1}} d^3p \,\, \left(p^i\right)^2 \left|\psi(\mathbf{p})\right|^2}    \,=\,  \frac{\beta}{4 \int\limits_{\widetilde{\mathbf{p}}^2 < 1} d^3\widetilde{p} \,\, \left(\widetilde{p}^i\right)^2\left|\widetilde{\psi}(\widetilde{\mathbf{p}})\right|^2} ~, ~~  \nonumber 
\end{eqnarray} where in the last expression we have used dimensionless quantities $\widetilde{p}^i = \beta^{1/2}p, \, \widetilde{\psi} = \beta^{-3/4}\psi$. From the normalization condition $\int\limits_{\widetilde{\mathbf{p}}^2 < 1} d^3\widetilde{p} \,\, \left|\widetilde{\psi}\right|^2 = 1$, one infers 

\begin{eqnarray} \int\limits_{\widetilde{\mathbf{p}}^2 < 1} d^3\widetilde{p} \,\, \left(\widetilde{p}^i\right)^2\left|\widetilde{\psi}(\widetilde{\mathbf{p}})\right|^2  \, \leq \, \int\limits_{\widetilde{\mathbf{p}}^2 < 1} d^3\widetilde{p} \,\, \left(\widetilde{p}^i\right)^2 \,=\,  \int\limits_{\widetilde{\mathbf{p}}^2 < 1} d^3\widetilde{p} \,\, \frac{\widetilde{\mathbf{p}}^2}{3} \,=\, \frac{4\pi}{15}~,  \nonumber \end{eqnarray} that concludes the proof.

To summarize the formalism, the quantum mechanics gets modified in the following way. For finding the energy spectrum we have now to solve the equation  

\begin{eqnarray}\label{Hamiltonian} \widehat{\mathcal{H}}\phi_n(\mathbf{x}) \,=\, \left[\frac{\widehat{\mathbf{P}}^2}{2m} \,+\, V\left(\widehat{\mathbf{X}}\right) \right] \phi_n(\mathbf{x}) \,=\,  \left[\frac{\widehat{\mathbf{p}}^2}{2m\left(1-\beta \widehat{\mathbf{p}}^2\right)^2}  \,+\, V\left(\mathbf{x}\right) \right] \phi_n(\mathbf{x}) \,=\, E_n\phi_n(\mathbf{x}) ~,  \end{eqnarray} 
provided that $\phi_n(\mathbf{x})$ has the Fourier representation 

\begin{equation}\label{cutoffFourierrep} \phi_n(\mathbf{x}) \,=\, \frac{1}{\left(2\pi\right)^{3/2}} \int\limits_{\mathbf{p}^2 < \beta^{-1}} d^3p \,\, e^{-i\mathbf{p}\cdot\mathbf{x}} \psi_n(\mathbf{p})~. \end{equation}

Because the complexity of the minimum-length modified Schr\"odinger equation, one might find it reasonable to forget some details and try to analyze it in the perturbative way. Namely, to rewrite Hamiltonian in the form

\begin{eqnarray}\label{pirvelimiakhvarianti}
\widehat{\mathcal{H}}  \,=\,  V\left(\mathbf{x}\right) \,+\, \sum\limits_{n =0}^{N} \frac{1+n}{2m} \, \beta^n \widehat{\mathbf{p}}^{2(n+1)} \,+\,O\left(\beta^{N+1}\right)~,
\end{eqnarray} and solve the equation $\mathcal{H}\phi_n(\mathbf{x}) = E_n\phi_n(\mathbf{x})$ order by order in $\beta$. Let us make the following generic comment about the using of standard perturbation method for solving this sort of equation. To the first order in $\beta$ one finds 

\begin{eqnarray}\label{corrham} \widehat{\mathcal{H}}  \,=\, \frac{\widehat{\mathbf{p}}^2}{2m} \,+\, V\left(\mathbf{x}\right) \,+\, \frac{\beta}{m} \,\widehat{\mathbf{p}}^4   ~, \end{eqnarray} for which the standard perturbation theory can be used for estimating leading order corrections to the energy and the wave-function \cite{LL3}   

\begin{eqnarray}\label{first} \widehat{\mathcal{H}}_0 \phi^{(0)}_k \,=\, E^{(0)}_k\phi^{(0)}_k ~,~~\delta E_k \,=\, \left\langle \phi^{(0)}_k\right|\widehat{\mathcal{H}}_1  \left| \phi^{(0)}_k \right\rangle   ~, ~~ \delta \phi_k   \,=\, \sum\limits_{j\neq k} \, \phi^{(0)}_j  \, \frac{\left\langle \phi^{(0)}_j\right|\widehat{\mathcal{H}}_1  \left| \phi^{(0)}_k \right\rangle}{E^{(0)}_k \,-\, E^{(0)}_j } ~,~~ \end{eqnarray} here $\widehat{\mathcal{H}}_1$ stands for the perturbation Hamiltonian $\widehat{\mathcal{H}}_1= \beta\widehat{\mathbf{p}}^4/m$. The Eq.(\ref{first}) can be considered to be good approximation for solving the equation $\widehat{\mathcal{H}} \phi_k \,=\, E_k\phi_k$ if certain conditions are satisfied. For clarifying this point let us recall what are the validity conditions for using WKB approximation. The WKB method is widely used in physics for obtaining an approximate solution for the linear differential equation whose highest derivative is multiplied by a small parameter \cite{LL3, BO}. Working in coordinate representation (for simplicity we consider one dimensional case given by Eq.\eqref{defxopo}) and keeping only leading correction, the equation $ \widehat{\mathcal{H}} \phi_k \,=\, E_k\phi_k $ turns into a fourth order differential equation  

\begin{eqnarray}\label{maineq} \frac{\beta}{3m} \, \frac{d^4\phi}{dx^4} \,-\, \frac{1}{2m}\, \frac{d^2\phi}{dx^2}  \,=\, \left[ E \,-\, V(x) \right] \phi~. \nonumber \end{eqnarray} The validity of the above approximation assumes that 

\begin{eqnarray}\label{validcon} \beta \left|\frac{d^4\phi^{(0)}}{dx^4}\right| \,\ll \, \left|\frac{d^2\phi^{(0)}}{dx^2}\right|~,  ~~\text{that is, }~~ \beta \left|\frac{d^4\phi^{(0)}}{dx^4}\right| \,\ll \,  \left|\left[  E \,-\, V(x)  \right] \phi^{(0)} \right|  ~. \end{eqnarray} Evidently, the condition \eqref{validcon} is violated near the turning points: $E \,-\, V(x) =0$.

The other "simplified" approach can be the following naive inversion of Eq.\eqref{Hamiltonian}

\begin{eqnarray}\label{meoremiakhvarianti}
\widehat{\mathbf{p}}^2\phi_n(\mathbf{x})   \,+\, 2m\left(1-\beta \widehat{\mathbf{p}}^2\right)^2\left[V\left(\mathbf{x}\right)\,-\, E_n \right] \phi_n(\mathbf{x}) \,=\,0~. ~~
\end{eqnarray} One immediately notices an important qualitative difference between Eqs.(\ref{pirvelimiakhvarianti}, \ref{meoremiakhvarianti}) that the "exact" equation \eqref{meoremiakhvarianti} is the 4-th order differential equation while the Eq.\eqref{pirvelimiakhvarianti} implies $2(N+1)$-th order differential equation.

Recently ml-QM has been applied to studying what happens to wave-function singularities upon minimum-length deformed quantization \cite{Battisti:2007jd, Battisti:2007zg, Montani:2007vu, Battisti:2008rv, Bouaziz:2007gs, Bouaziz:2010hc}. An interesting question that occurs at this stage is to ask: What criteria constitute the singularity avoidance, is the singularity avoided in quantum dynamics? Throughout this paper we address this question. In what follows, by considering various examples (including the minimum-length deformed Wheeler-De Witt equation) we will demonstrate that only proper treatment given by Eqs.(\ref{Hamiltonian}, \ref{cutoffFourierrep}) leads to the avoidance of wave-function singularities in the unique way.

\section{How does ml-QM regularize singularities? A few examples}

{\bf Example 1. (Exact solution).} As a first illustrative example, let us consider the motion of a particle inside a spherical cavity of radius $R$ with impenetrable walls

\begin{eqnarray}
 V(r) \, = \,
   \begin{cases}
      0 \, ,     &\text{if ~~ $ r< R  $~,}\\
      \infty \,,          &\text{if ~~\, $r \geq R $~.}
   \end{cases} 
\nonumber \end{eqnarray} Recalling that the radial part of the Laplace operator in spherical coordinates has the form 

\[ \Delta \psi \,=\, \frac{1}{r^2} \, \frac{\partial}{\partial r} \left( r^2\frac{\partial \psi}{\partial r} \right) ~, \] after making the substitution $\psi = \chi /r $ one finds \[ \Delta \frac{\chi}{r} = \frac{1}{r}\, \frac{\partial^2\chi}{\partial r^2} ~,~\Delta^2  \frac{\chi}{r} =  \frac{1}{r}\, \frac{\partial^4\chi}{\partial r^4} ~, ~ \Delta^n  \frac{\chi}{r} =  \frac{1}{r}\, \frac{\partial^{2n}\chi}{\partial r^{2n}}~.  \] For $\ell =0$ we seek the eigenfunctions of the Hamiltonian inside the cavity   

\begin{eqnarray} \frac{ \widehat{\mathbf{p}}_r^2 }{2m\left(1-\beta \widehat{\mathbf{p}}_r^2\right)^2} \, \psi(r) \,=\, E\psi(r) \nonumber \end{eqnarray} in the form $\psi \propto \exp(i k r) /r$. So, we find 

\[ \frac{ k^2 }{\left(1-\beta k^2\right)^2}  \,=\, 2mE ~, \] the eigenfunctions can be written in the standard form

\begin{eqnarray} \chi_n \,=\,  \sqrt{\frac{2}{R}} \sin \left(\frac{n\pi r}{R} \right) ~, ~~(n \,=\, 1,\,2,\,3 \ldots) ~,  \nonumber \end{eqnarray} that results in the energy spectrum   

\begin{eqnarray} k_n \,=\, \frac{n\pi}{R}~, ~~\Rightarrow ~~   E_n \,=\,  \frac{ k_n^2 }{2m\left(1-\beta k_n^2\right)^2} ~.  \nonumber  \end{eqnarray}

\noindent Because of the cut-off $k_n < \beta^{-1/2}$ we can not take the limit $R\rightarrow 0$, the radius of the cavity is now bounded from below by the condition $R > n\pi \sqrt{\beta}$. One can estimate the pressure exerted on the walls of cavity by using the first law of thermodynamics $dE = - \mathcal{P}dV$

\begin{eqnarray} \mathcal{P}_n \,=\, -\, \frac{\partial E_n}{\partial R} \, \frac{1}{4\pi R^2} \,=\,   \frac{n}{4mR^4}\left[   \frac{ k_n }{\left(1-\beta k_n^2\right)^2} \,+\, \frac{ 2\beta k_n^3 }{\left(1-\beta k_n^2\right)^3} \right]   ~. \nonumber  \end{eqnarray} So, the pressure in the $\psi_n$ state becomes infinite as $R$ approaches the value $n\pi \sqrt{\beta}$, or otherwise, when $k_n$ approaches $\beta^{-1/2}$.

{\bf Example 2. (Semi-classical analysis).} Let us now consider a semi-classical treatment of the particle motion in the potential\footnote{One may think of it as a Coulomb potential in a higher-dimensional space.} $-\alpha / r^s$ by using the Hamiltonian \eqref{Hamiltonian}. A particle confined to a small region of
radius $r$ about the origin will have (because of quantum fluctuations) the momentum of the order of $1/r$ (that is estimated simply via the uncertainty relation\footnote{Let us recall that between $x_i$ and $p_k$ there are standard uncertainty relations: $\delta x_i\delta p_k =\delta_{ik}/2$.}). The average value of the energy is approximately \cite{LL3, Krainov:1977kg, Weisskopf:1970jx}

\begin{eqnarray}\label{semiclassical} E(r) \,=\, \frac{\mathbf{p}^2} {2m \left(1 \,-\, \beta \mathbf{p}^2 \right)^2 } \,-\,  \frac{\alpha}{r^s} \,\simeq \,  \frac{r^2}{m\left(r^2 \,-\,\beta \right)^2} \,-\,  \frac{\alpha}{r^s} ~. \end{eqnarray} It is evident that even for $s>2$ there is lower bound on the energy. Namely, when $r$ approaches $\beta^{1/2}$ the first term in Eq.\eqref{semiclassical} becomes sharply growing thus compensating the second term. If we restrict ourselves to the leading correction in $\beta$, the Eq.\eqref{semiclassical} takes the form 

\begin{eqnarray}\label{leadingcorrenergy}&& E(r) \,\simeq \, \frac{1}{m r^2 } \,+\, \frac{2\beta}{r^4} \,-\,  \frac{\alpha}{r^s} ~. \end{eqnarray} From eq.\eqref{leadingcorrenergy} one infers that the fall to the center can be avoided if $s<4$. Thus, the inclusion of higher and higher corrections in $\beta$ will stabilize the wave-function "collapse" for higher and higher values of $s$, but it is the exact expression \eqref{semiclassical} that shows that the fall to the center is avoided for an arbitrary value of $s$. 

{\bf Example 3. (Perturbative treatment).} Going beyond the semi-classical treatment, let us first notice that for strongly attractive potentials the wave function oscillates indefinitely on the way to the origin, that is, there is a divergence as $r\rightarrow 0$ producing an infinite oscillation representing
the classical singularity \cite{Case:1950an, Perelomov:1970fz, Frank:1971xx}. Now let us use the perturbative treatment and see how the $\beta$ term in Eq.\eqref{corrham} might alter the situation. After the substitution $\phi = \psi/r$, the radial part of the Schr\"odinger equation for $\ell =0$ takes the form

\begin{eqnarray}\label{radialeqzerol} \beta \frac{d^4\psi}{dr^4} \,+\, \left[  \frac{4\beta}{r^2} \,-\, \frac{1}{2}\right] \frac{d^2\psi}{dr^2} \,-\, \frac{8\beta}{r^3} \frac{d\psi}{dr} \,+\,  \left[  \frac{8\beta}{r^4} \,-\, \frac{1}{ r^2} \,-\, mE \,-\, \frac{m\alpha}{ r^{s}}\right]
\psi(r) \,=\, 0~.~~ \end{eqnarray} Assuming $0<s<4$, then for $r \ll \sqrt{\beta}$ the Eq.\eqref{radialeqzerol} is completely dominated by the $\beta$ terms and reduces to 

\begin{eqnarray}  \frac{d^4\psi}{dr^4} \,+\,  \frac{4}{r^2} \, \frac{d^2\psi}{dr^2} \,-\, \frac{8}{r^3} \, \frac{d\psi}{dr}   \,+\, \frac{8}{r^4} \, \psi \,=\, 0 ~, \nonumber \end{eqnarray} for which one readily finds the following solution: $\psi  \sim r^{\nu}$   

\begin{eqnarray}&& \nu^4 \,-\, 6\nu^3 \,+\, 15\nu^2 \,-\, 18\nu \,+\, 8 \,=\, 0 ~, ~~\nu_1=1~ ,~ \nu_2=2 ~,~ \nu_3=\frac{3-i\sqrt
{7}}{2}~,~\nu_4=\frac{3+i\sqrt{7}}{2}  ~, \nonumber \\&&
\phi(r) \,=\, A_1 \,+\, A_2 r \,+\,  A_3\sqrt{r}\cos\left(\frac{\sqrt{7}}{2}\ln\frac{r}{\beta}\right)+A_4\sqrt{r}\sin\left(\frac{\sqrt{7}}{2}\ln\frac{r}{\beta}\right) ~.
\label{smalregsolution}\end{eqnarray} The singular behaviour is reflected by the last two terms in Eq.\eqref{smalregsolution}. So, to avoid them we have to take $A_3=A_4 =0$. For distance $r \gg \sqrt{\beta}$ the beta terms become less essential; in this regime one arrives at the standard Schr\"odinger equation. Thus, roughly speaking, the solution for $r\lesssim \sqrt{\beta}$ will be given by Eq.\eqref{smalregsolution} with $A_3=A_4 =0$ that should be matched with the solution of the standard Schr\"odinger equation (representing the solution for $r \gtrsim \sqrt{\beta}$) in the vicinity of $r\simeq \sqrt{\beta}$.

\section{Cosmological singularities in light of the minimum-length deformed Wheeler-De Witt equation} 
 
\subsubsection{Wheeler-De Witt equation }

In the spirit of the above discussion one can address the problem of cosmological singularity as well.

\[\mathcal{L} \,=\,  \sqrt{-g} \left[ - \, \frac{R}{16\pi G_N} \,+\,  \frac{\partial_{\mu} \phi \partial^{\mu}\phi}{2}  \,-\, U\left(\phi\right) \right] ~,  \] in a closed Friedmann-Lema\^{i}tre-Robertson-Walker metric

\begin{eqnarray} ds^2 \,=\, N^2(t)dt^2 \,-\, a^2(t) \left[\frac{dr^2}{1-\kappa r^2} \,+\, r^2 d\Omega_2^2 \right]  \,=\,  N^2(t)dt^2 \,-\, b^2(t)d\Omega_3^2  ~,  \nonumber \end{eqnarray} where $d\Omega_2^2$ and $d\Omega_3^2$ are the line elements of the unit 2 and 3 spheres, respectively, and $b = a\kappa^{-1/2}$. So, the scale factor $b$ has the dimension of length. The Wheeler-DeWitt equation takes the form \cite{Linde:2005ht}

\begin{eqnarray}\label{WheelerDeWitt} \left[ \frac{G_Nb^{-q}  \widehat{p}_b \, b^q\widehat{p}_b}{3\pi}    \,-\, \frac{\widehat{p}_{\phi}^2}{4\pi^2 b^2}   \,+\, \frac{3\pi b^2}{4G_N}  \,-\, 2\pi^2 b^4 U\left(\phi\right)  \right] \Psi(b,\,\phi) \,=\, 0~. \end{eqnarray} 

\noindent  where $\widehat{p}_b = -i\partial/\partial b$ and $\widehat{p}_{\phi} = -i\partial/\partial \phi$ and the combination $b^{-q}  \widehat{p}_b \, b^q\widehat{p}_b$ accounts for the operator ordering ambiguity. Let us for simplicity consider a specific case $U=0$. In this case Eq.\eqref{WheelerDeWitt} reduces to    

\begin{eqnarray}\label{gantolebawd} \left[ \frac{G_Nb^{-q}  \widehat{p}_b \, b^q\widehat{p}_b}{3\pi}    \,-\, \frac{\widehat{p}_{\phi}^2}{4\pi^2 b^2}   \,+\, \frac{3\pi b^2}{4G_N}   \right] \Psi(b,\,\phi) \,=\, 0~, \nonumber \end{eqnarray} where now variables $b$ and $\phi$ can be separated $\Psi(b,\,\phi) = \Psi(b)\exp(i\sigma \phi)$ leading to 

\begin{eqnarray}\label{WheelerDeWittsimplified} \left[ \frac{G_Nb^{-q}  \widehat{p}_b \, b^q\widehat{p}_b}{3\pi}    \,-\, \frac{\sigma^2}{4\pi^2 b^2}   \,+\, \frac{3\pi b^2}{4G_N}   \right] \Psi(b) \,=\, 0~.  \end{eqnarray}

\subsubsection{Minimum-length modified Wheeler-De Witt equation }

Let us forget about the operator ordering (we take $q=0$) and write minimum-length modified version of Eq.\eqref{gantolebawd} in the following form

\begin{eqnarray}\label{modifitsirebuligant}&& \left[ \frac{G_N  \widehat{P}_b^2 }{3\pi}    \,-\, \frac{\widehat{P}_{\phi}^2}{4\pi^2 b^2}   \,+\, \frac{3\pi b^2}{4G_N}   \right] \Psi(b,\,\phi) \,=\, \nonumber \\&& \left[ \frac{G_N  \widehat{p}_b^2 }{3\pi \left[ 1 - \beta \left( \widehat{p}_b^2 +\widehat{p}_{\phi}^2 \right)\right]^2}    \,-\, \frac{\widehat{p}_{\phi}^2}{4\pi^2 b^2 \left[ 1 - \beta \left( \widehat{p}_b^2 +\widehat{p}_{\phi}^2 \right)\right]^2}   \,+\, \frac{3\pi b^2}{4G_N}   \right] \Psi(b,\,\phi) \,=\,  0~.  \nonumber \end{eqnarray} Making substitution $\Psi(b,\,\phi) = \Psi(b)\exp(i\sigma \phi)$ this equation reduces to

\begin{eqnarray} \left[ \frac{G_N  \widehat{p}_b^2 }{3\pi \left[ 1 - \beta \left( \widehat{p}_b^2 +\sigma^2 \right)\right]^2}    \,-\,  \frac{\sigma^2}{4\pi^2 b^2 \left[ 1 - \beta \left( \widehat{p}_b^2 +\sigma^2 \right)\right]^2}   \,+\, \frac{3\pi b^2}{4G_N}   \right] \Psi(b) \,=\,  0~.  \label{ertertibologant} \end{eqnarray}

\noindent Denoting

\[ \widetilde{\Psi}(b) \,\equiv\,  \frac{1}{ \left[ 1 - \beta \left( \widehat{p}_b^2 +\sigma^2 \right)\right]^2} \, \Psi(b)~,  \] the Eq.\eqref{ertertibologant} takes the form (actually, here we use the inversion prescription Eq.\eqref{meoremiakhvarianti})

\begin{eqnarray} \left( \frac{G_N  \widehat{p}_b^2 }{3\pi }    \,-\, \frac{\sigma^2}{4\pi^2 b^2 }   \,+\, \frac{3\pi b^2}{4G_N}  \left[ 1 - \beta \left( \widehat{p}_b^2 +\sigma^2 \right)\right]^2 \right) \widetilde{\Psi}(b)  \,=\,  0~,  \nonumber  \end{eqnarray} which in the limit $b \rightarrow 0$ can be written in a somewhat simplified form

\begin{eqnarray} \left( \frac{G_N  \widehat{p}_b^2 }{3\pi }   \,+\, \frac{3\pi b^2 \widehat{p}_b^4}{4G_N}   \,-\, \frac{\sigma^2}{4\pi^2 b^2 }  \right) \widetilde{\Psi}(b)  \,=\,  0~. \label{gamartivebuligantoleba}  \end{eqnarray} From Eq.\eqref{gamartivebuligantoleba} one sees that if this equation admits a regular solution $ \Psi(b)$ it should occur at the expense of $\widehat{p}_b^4$ term. Looking at different solutions of equation 

 \[  \left(\frac{d^4}{dz^4} \,-\, \frac{\varkappa
}{z^4} \right)     \widetilde{\Psi}(z) \,=\, 0~,   \] (we use the package Mathematica-8)

\begin{eqnarray}&& \varkappa = 0.1~,~~ \widetilde{\Psi}(z) = \frac{C_1}{z^{0.0161823}} + C_2 \times z^{1.05146}  +  C_3 \times z^{1.94854}  + C_4\times  z^{3.01618} ~, \\&& \varkappa = 1~,~~ \widetilde{\Psi}(z) = C_1\times z^{\left.\left(3- \sqrt{5 + 4 \sqrt{2}}\right)\right/2}  + 
 C_2\times z^{\left.\left(3+ \sqrt{5 + 4 \sqrt{2}}\right)\right/2}  + 
  \nonumber \\&&  C_3\times z^{3/2}  \cos\left(\frac{\sqrt{-5 + 4 \sqrt{2}}}{2} \ln(z)\right) + C_4\times z^{3/2}  \sin\left(\frac{\sqrt{-5 + 4 \sqrt{2}}}{2} \ln(z)\right) ~, \\&& \varkappa = 50~,~~ \widetilde{\Psi}(z) = C_1\times z^{\left.\left(3- \sqrt{5 + 4 \sqrt{51}}\right)\right/2}  + 
 C_2\times z^{\left.\left(3+ \sqrt{5 + 4 \sqrt{51}}\right)\right/2}  +  \nonumber \\&&
 C_3\times z^{3/2}  \cos\left(\frac{\sqrt{-5 + 4 \sqrt{51}}}{2} \ln(z)\right)  + C_4\times z^{3/2}  \sin\left(\frac{\sqrt{-5 + 4 \sqrt{51}}}{2} \ln(z)\right) ~,
\end{eqnarray} one concludes that the wave-function $\Psi(b)  \,=\, \left[ 1 - \beta \left( \widehat{p}_b^2 +\sigma^2 \right)\right]^2 \widetilde{\Psi}(b) $, which contains the fourth derivative of $\widetilde{\Psi}(b)$, can not be regular at $b=0$ for arbitrary values of $\sigma$.

Having discussed the above subtleties of the wave function in ml-QM, we can now enquire into the interpretation of it.    

\section{The issues of wave function and gauge invariance in ml-QM}

The Lagrangian for a non-relativistic charged particle moving in a given electromagnetic field is given by \cite{LL2}-\S 16

\begin{eqnarray}\label{ararelnatselmagveli}  \mathcal{L} \,=\,  \frac{m\dot{\mathbf{r}}^2}{2} \,+\, q \mathbf{A}\cdot\dot{\mathbf{r}} \,-\, q \phi  ~, \end{eqnarray} and hence the Hamiltonian takes the form, $\mathbf{P} = m\dot{\mathbf{r}} + q \mathbf{A}$, 

\begin{eqnarray}\label{ararelnatsurtgareveltan}  \mathcal{H} \,=\,  \frac{\left( \mathbf{P} \,-\,q \mathbf{A}  \right)^2}{2m}  \,+\, q \phi  ~. \end{eqnarray} The transition to quantum mechanics implies the replacement of $\mathbf{P}$ with the momentum operator \cite{LL3}-\S 111. Following this prescription, in the case of ml-QM one obtains         

\begin{eqnarray}\label{Hamiltonianem}  \widehat{\mathcal{H}} \,=\,  \frac{\left( \widehat{\mathbf{P}} \,-\,q \mathbf{A}  \right)^2}{2m}  \,+\, q \phi \,=\,   \frac{\widehat{\mathbf{P}}^2}{2m} \,-\, \frac{q}{2m} \left( \widehat{\mathbf{P}}\cdot\mathbf{A}  \,+\,   \mathbf{A}\cdot\widehat{\mathbf{P}} \right) \,+\,\frac{q^2\mathbf{A}^2}{2m}  \,+\, q \phi   ~. \end{eqnarray}

\noindent Let us now derive the quantum mechanical expression for the electric current. It can be derived via the formula

\begin{equation}\label{LLgamosakhulebakmdenistvis}  \delta \langle H \rangle_{\Psi} \,=\, - \, \int d^3x \, \delta \mathbf{A} \cdot  \mathbf{J}  ~, \end{equation} where $ \delta \langle H \rangle_{\Psi} $ stands for the variation of $\langle \Psi |\widehat{\mathcal{H}}  | \Psi \rangle$ caused by the variation $\delta \mathbf{A}$, see \cite{LL3}-\S 115. The variation of Eq.\eqref{Hamiltonianem} with respect to $\delta \mathbf{A}$ amounts to the equation

\begin{eqnarray}  \delta \langle \Psi |\widehat{\mathcal{H}}  | \Psi \rangle \,=\,  \int d^3x \, \Psi^*\left[ -\frac{q\left(\widehat{\mathbf{P}}\cdot \delta\mathbf{A} \,+\, \delta\mathbf{A} \cdot \widehat{\mathbf{P}}  \right)}{2m}   \,+\,\frac{q^2\mathbf{A}\cdot \delta\mathbf{A}}{m}  \right] \Psi   ~, \nonumber  \end{eqnarray} which after using the equality 

\begin{equation}  \int d^3x \, \Psi^*  \widehat{\mathbf{P}}\cdot \delta\mathbf{A}  \Psi  \,=\, -  \int d^3x \, \delta\mathbf{A} \cdot  \Psi \widehat{\mathbf{P}}\Psi^*     ~, \nonumber \end{equation} (that can easily be obtained by using repeated integration and taking into account that each time the surface integral at spatial infinity vanishes) reduces to

\begin{eqnarray}\label{variationHamiltonian}  \delta \langle \Psi |\widehat{\mathcal{H}}  | \Psi \rangle \,=\, \int d^3x \, \frac{q^2\mathbf{A}\cdot \delta\mathbf{A}}{m} \, \Psi^* \Psi  \,+\,   \int d^3x \, \frac{q \delta\mathbf{A} \cdot \left( \Psi \widehat{\mathbf{P}}\Psi^* \,-\, \Psi^* \widehat{\mathbf{P}}\Psi  \right)}{2m}    ~.  \end{eqnarray} In absence of the background field, from Eq.\eqref{variationHamiltonian} one finds the current density in the form

\begin{eqnarray}\label{kvantmechdeni} \mathbf{J} \,=\, \frac{\Psi^* \widehat{\mathbf{P}}\Psi  \,-\, \Psi \widehat{\mathbf{P}}\Psi^*}{2m} \,=\,  \frac{i}{2m} \left( \Psi \frac{\nabla}{1 \,+\, \beta \Delta}\Psi^*  \,-\,\Psi^* \frac{\nabla}{1 \,+\, \beta \Delta}\Psi  \right) \,  \, ~,   \end{eqnarray} which clearly indicates that the continuity equation

\[ \frac{\partial \left(\Psi^*\Psi\right)}{\partial t} \,+\, \mbox{div}  \mathbf{J} \,=\, 0~, \] does not hold any more.

The expression of current that would satisfy the continuity equation immediately follows from the equation

\begin{eqnarray}\label{shenakhvadidenischrgantolebidan}
2mi\partial_t(\Psi^*\Psi) = \Psi^*\widehat{\mathbf{P}}^2\Psi - \Psi \widehat{\mathbf{P}}^2\Psi^* = \sum\limits_{n =0}^{\infty} (1+n)\beta^n  \left[\Psi^*\left(-\Delta\right)^{(n+1)}\Psi - \Psi \left(-\Delta\right)^{(n+1)}\Psi^*  \right]   ~. ~~~~
\end{eqnarray}

Using the partial-differentiation 
\begin{eqnarray}&& \Psi^*\partial^2\Psi = \partial( \Psi^*\partial\Psi) -  \partial\Psi^*\partial\Psi~, \nonumber \\&& \Psi^*\partial^4\Psi = \partial( \Psi^*\partial^3\Psi) - \partial\Psi^*\partial^3\Psi =  \partial( \Psi^*\partial^3\Psi) -  \partial(\partial\Psi^*\partial^2\Psi)+ \partial^2\Psi^*\partial^2\Psi ~, \nonumber \\&& \Psi^*\partial^6\Psi = \partial( \Psi^*\partial^5\Psi) - \partial\Psi^*\partial^5\Psi =  \partial( \Psi^*\partial^5\Psi) - \partial(\partial\Psi^*\partial^4\Psi) + \partial^2\Psi^*\partial^4\Psi =  \nonumber \\&& \partial( \Psi^*\partial^5\Psi) - \partial(\partial\Psi^*\partial^4\Psi) + \partial(\partial^2\Psi^*\partial^3\Psi) - \partial^3\Psi^*\partial^3\Psi ~. \nonumber \end{eqnarray} one finds

\begin{eqnarray}\label{shenakhvadikvanmechdeni}
&&\sum\limits_{n =0}^{\infty} (1+n)\beta^n  \left[\Psi^*\left(-\Delta\right)^{(n+1)}\Psi \,-\, \Psi \left(-\Delta\right)^{(n+1)}\Psi^*  \right]  \,=\,  \nonumber\\&& \text{div} \left[ \, \sum\limits_{n =0}^{\infty} (1+n)\beta^n (-1)^{n+1} \left(\Psi^*\nabla^{2n+1}\Psi \,-\, \Psi \nabla^{2n+1}\Psi^*  \right) \,+\, \right. \nonumber\\&&    \sum\limits_{n =1}^{\infty} (1+n)\beta^n (-1)^{n+1} \left(\nabla\Psi^*\nabla^{2n}\Psi \,-\, \nabla\Psi \nabla^{2n}\Psi^*  \right) \,+\,   \nonumber\\&&  \sum\limits_{n =2}^{\infty} (1+n)\beta^n (-1)^{n+1} \left(\nabla^2\Psi^*\nabla^{2n-1}\Psi \,-\, \nabla^2\Psi \nabla^{2n-1}\Psi^*  \right)  \,+\,  \nonumber\\&& \left.   \sum\limits_{n =3}^{\infty} (1+n)\beta^n (-1)^{n+1} \left(\nabla^3\Psi^*\nabla^{2n-2}\Psi \,-\, \nabla^3\Psi \nabla^{2n-2}\Psi^*  \right)  \,+\, \cdots ~~~~~~~~~~~~~ \right]   ~. ~~\end{eqnarray}

\noindent Let us notice that this sort of current (truncated to first order in $\beta$) is used in \cite{Stetsko:2007ze}.

Another interesting point is the gauge invariance. In the framework of quantum mechanics, the gauge transformation $\mathbf{A} \rightarrow \mathbf{A} + \nabla f,\, \phi \rightarrow \phi -\partial_tf$ in the Schr\"{o}dinger equation is compensated by the wave-function transformation $\Psi \rightarrow \Psi\exp(iqf)$. But now this sort of gauge invariance is violated. It implies that the vector potential has to be seen as a real physical field in ml-QM \cite{Ehrenberg}. This point will be addressed in more detail in the following section.

\section{Interaction of charged field with the electromagnetic one }

The action for a charged scalar field in the framework of ml-QFT can be written in the form

\begin{eqnarray}\label{damukhtscalqmedebamod}  \mathcal{W}[\varPhi] = - \int \frac{d^4x}{2} \left[\widehat{P}_{\mu}\varPhi \widehat{P}^{\mu}\varPhi^* 
+ m^2\varPhi\varPhi^* \right] =   - \int \frac{d^4x}{2} \left[\widehat{P}_{0}\varPhi \widehat{P}^{0}\varPhi^* -  \widehat{\mathbf{P}}\varPhi \cdot \widehat{\mathbf{P}}\varPhi^*
+ m^2\varPhi\varPhi^* \right]  ~,~~~~ \end{eqnarray} where $\widehat{P}_0 = i\partial_t$. That is, ($\mathcal{D}
_{t} \equiv \partial_t,\, \mathcal{D}
_{i} \equiv \partial_i (1+\beta\Delta)^{-1}$)

\begin{eqnarray}\label{scactionone}  \mathcal{W}[\varPhi] =   \int \frac{d^4x}{2}  \left[\mathcal{D}
_{\mu}\varPhi \mathcal{D}
^{\mu}\varPhi^* \,-\, m^2\varPhi\varPhi^* \right]  ~, \end{eqnarray} which results in the equation of motion 

\begin{equation}\label{modzraobisgantoleba} \mathcal{D}
_{\mu}\mathcal{D}
^{\mu} \,\varPhi \,+\, m^2\varPhi  \,\equiv\, \partial_{t}^2\varPhi \,-\, \frac{\Delta}{\left(1+\beta \Delta\right)^2}\,\varPhi \,+\, m^2\varPhi  \,=\, 0~. \nonumber \end{equation} 

\subsection{ml-QFT - Noether current } 

Clearly, the Lagrangian has the $U(1)$ symmetry $\varPhi \rightarrow e^{i\alpha}\varPhi,\, \varPhi^* \rightarrow e^{-i\alpha}\varPhi^*$, which (in the standard case) is connected with the conservation of electromagnetic current. Following the reasoning of Noether's theorem one finds ($\varPhi_1\equiv \varPhi,\,\varPhi_2\equiv\varPhi^*$)

\begin{eqnarray}\label{shualedgant} \frac{\partial \mathcal{L}}{\partial \alpha} = 0 = \frac{\partial\mathcal{L}}{\partial \varPhi_k} \, \frac{\partial \varPhi_k}{\partial\alpha} + \frac{\partial\mathcal{L}}{\partial\left(\mathcal{D}
_{\mu}\varPhi_k\right)} \, \frac{\partial \left(\mathcal{D}
_{\mu} \varPhi_k \right)}{\partial\alpha} =  \frac{\partial\mathcal{L}}{\partial \varPhi_k} \, \frac{\partial \varPhi_k}{\partial\alpha} + \frac{\partial\mathcal{L}}{\partial\left(\mathcal{D}
_{\mu}\varPhi_k\right)} \, \mathcal{D}
_{\mu} \frac{\partial \varPhi_k}{\partial\alpha} ~.\end{eqnarray} Using here the equation of motion 

\begin{eqnarray} \frac{\partial\mathcal{L}}{\partial \varPhi_k}  \,=\, \mathcal{D}_{\mu} \frac{\partial\mathcal{L}}{\partial\left(\mathcal{D}_{\mu}\varPhi_k\right)} ~, \nonumber \end{eqnarray} from Eq.\eqref{shualedgant} one gets

\begin{eqnarray}\label{shualedurigantoleba}&& \left(\mathcal{D}_{\mu} \frac{\partial\mathcal{L}}{\partial\left(\mathcal{D}_{\mu}\varPhi_k\right)} \right) \frac{\partial \varPhi_k}{\partial\alpha} + \frac{\partial\mathcal{L}}{\partial\left(\mathcal{D}_{\mu}\varPhi_k\right)} \, \mathcal{D}_{\mu} \frac{\partial \varPhi_k}{\partial\alpha} =  \nonumber\\&& ~~~~~~~~~~~~~~~~~~~~~~ i\left(\mathcal{D}_{\mu} \mathcal{D}^{\mu}\varPhi^* \right)  \varPhi - i\left(\mathcal{D}_{\mu} \mathcal{D}^{\mu}\varPhi \right)  \varPhi^* + i\mathcal{D}^{\mu}\varPhi^*   \mathcal{D}_{\mu} \varPhi -i\mathcal{D}^{\mu}\varPhi   \mathcal{D}_{\mu} \varPhi^*  =0 ~~.\end{eqnarray} The equation \eqref{shualedurigantoleba} reduces to

\begin{eqnarray}&& \partial_{\mu}\left[i\left(\partial^{\mu}\varPhi^*\right)\varPhi -i\varPhi^*\partial^{\mu}\varPhi  \right] -i \sum\limits_{n =1}^{\infty} (1+n)\beta^n  \left[\varPhi^*\left(-\Delta\right)^{(n+1)}\varPhi \,-\, \varPhi \left(-\Delta\right)^{(n+1)}\varPhi^*  \right] - \nonumber\\&&  i \partial_i \varPhi^*  \sum\limits_{l=1}^{\infty}\partial_i\left(-\beta\Delta\right)^l \varPhi +  i\sum\limits_{n=1}^{\infty}\partial_i\left(-\beta\Delta\right)^n\varPhi  \partial_i \varPhi^* -  i \sum\limits_{n=1}^{\infty}\partial_i\left(-\beta\Delta\right)^n \varPhi^*  \sum\limits_{l=1}^{\infty}\partial_i\left(-\beta\Delta\right)^l \varPhi + \nonumber\\&& i\sum\limits_{n=1}^{\infty}\partial_i\left(-\beta\Delta\right)^n\varPhi  \sum\limits_{l=1}^{\infty}\partial_i\left(-\beta\Delta\right)^l \varPhi^* =0 ~.\nonumber \end{eqnarray} To the first order in $\beta$ this relation takes the form

\begin{eqnarray}&& \partial_{\mu}\left[i\left(\partial^{\mu}\varPhi^*\right)\varPhi -i\varPhi^*\partial^{\mu}\varPhi  \right] -i \beta   \left[2\varPhi^*\Delta^{2}\varPhi \,-\, 2\varPhi \Delta^{2}\varPhi^*  - \partial_i \varPhi^*  \partial_i\Delta \varPhi + \partial_i\Delta\varPhi  \partial_i \varPhi^*\right]  \,=\, \nonumber \\&& \partial_{\mu}\left[i\left(\partial^{\mu}\varPhi^*\right)\varPhi -i\varPhi^*\partial^{\mu}\varPhi  \right] -i \beta  \partial_j \left[ 2\varPhi^*\partial_j\Delta\varPhi \,-\, 2\varPhi\partial_j\Delta\varPhi^* \,-\,3\partial_j \varPhi^*\Delta\varPhi \,+\,  \Delta\varPhi\partial_j\varPhi^*\right] =0 ~. \nonumber \end{eqnarray} So that the zero component of the current will have the standard form

\begin{equation}\label{denisnulovanikomponenta} J^0 \,=\, i\varPhi^*\partial^{0}\varPhi \,-\, i\left(\partial^{0}\varPhi^*\right)\varPhi ~, \end{equation} while $J^i$ gets modified as

\begin{equation}\label{denissivrtsulikomponenta} J^j \,=\, i\varPhi^*\partial^{j}\varPhi \,-\, i\left(\partial^{j}\varPhi^*\right)\varPhi -i \beta  \left[ 2\varPhi^*\partial^j\Delta\varPhi \,-\, 2\varPhi\partial^j\Delta\varPhi^* \,-\,3\partial^j \varPhi^*\Delta\varPhi \,+\,  \Delta\varPhi\partial^j\varPhi^*\right] ~.\end{equation}

\subsection{Coupling to the electromagnetic field}

The coupling to an external electromagnetic field can be introduced by replacing $P_{\mu}$ in Eq.\eqref{scactionone} with the $P_{\mu} -eA_{\mu}$ \cite{LL4}-\S 32. That means to define "covariant" derivative as 

\begin{equation}\label{elmagvelischartvistsesi}
\nabla_{\mu} \,=\, \mathcal{D}
_{\mu} \,+\, ieA_{\mu}~,
\end{equation}

\begin{eqnarray}\label{damukhtuliskalari} \mathcal{W}[\varPhi] = \int d^4x \, \frac{1}{2} \left[\mathcal{D}_{\mu}\varPhi \mathcal{D}^{\mu}\varPhi^* \,-\, m^2\varPhi\varPhi^*  \,+\,   ie A_{\mu} \left\{ \varPhi^*\mathcal{D}^{\mu}\varPhi \,-\,\left(\mathcal{D}^{\mu}\varPhi^*\right)\varPhi\right\} \right]  ~,\end{eqnarray}

\noindent Let us emphasize that the action \eqref{damukhtuliskalari} is not invariant under the gauge transformation 

\[ \varPhi \,\rightarrow\, e^{i\alpha}\varPhi~, ~~~ A_{\mu}  \,\rightarrow\,  A_{\mu} \,-\, \frac{ie^{-i\alpha}}{e} \, \mathcal{D}_{\mu}e^{i\alpha}~. \] This description of the electromagnetic coupling  seems to us the most intuitive picture following immediately from Eq.\eqref{Hamiltonianem} than making the gauging of $U(1)$ symmetry in Eq.\eqref{scactionone}, that is, replacing $\partial_{\mu}$ in Eq.\eqref{scactionone} with $\partial_{\mu} + ieA_{\mu}$ \cite{Hossenfelder:2003jz, Kober:2010sj}.

To have an uniform picture, the kinetic term for the electromagnetic field can be introduced merely by replacing: $\partial_{\mu} \rightarrow \mathcal{D}_{\mu}$ in the tensor of the electromagnetic field 

\[ F_{\mu\nu} \,=\, \partial_{\mu}A_{\nu} \,-\, \partial_{\nu}A_{\mu} \,\rightarrow \, \mathcal{F}_{\mu\nu} \,=\, \mathcal{D}_{\mu}A_{\nu} \,-\, \mathcal{D}_{\nu}A_{\mu} ~. \] This definition of kinetic term goes back to the paper \cite{Hossenfelder:2003jz}. For yet another approach to the electromagnetic kinetic term see \cite{Ashoorioon:2004rs}. An alternative path used in paper \cite{Ashoorioon:2004rs} consists of the following steps. The standard action is taken as a starting point

\begin{eqnarray} \mathcal{W} \,=\, -\, \frac{1}{4}\int d^4x \, F_{\mu\nu}F^{\mu\nu} \,=\,  \frac{1}{4}\int  d^4x \, \left[ 2F_{0i}F_{0i} \,-\, F_{ik}F_{ik} \right] \,=\,   \frac{1}{2}\int  d^4x \, \left[ \mathbf{E}^2 \,-\, \mathbf{B}^2 \right]  ~. \nonumber \end{eqnarray} Then the the gauge conditions $A_0 =\partial_iA_i =0$ are imposed that results in

\[ \mathcal{W} \,=\, \frac{1}{2}\int dt \, d^3x \, \left[ \left( \partial_0 \mathbf{A} \right)^2 \,-\, \left( \nabla\times \mathbf{A} \right)^2 \right]  ~. \] The operator $\nabla$ is identified with $\mathbf{p} = -i\nabla$ and the modification due to momentum deformation is understood as

\[ \mathcal{W} \,=\, \frac{1}{2}\int dt \, d^3x \, \left[ \left( \partial_0 \mathbf{A} \right)^2 \,-\, \left( \mathbf{P}\times \mathbf{A} \right)^2 \right]  ~. \]

In analogy to the $U(1)$ case, for the Yang-Mills field one can write 

\[\mathcal{W} \,=\, -\frac{1}{4}\int d^4x \, \text{Sp}\left(\mathcal{F}_{\mu\nu}\mathcal{F}^{\mu\nu} \right) ~ , ~~~~  \mathcal{F}_{\mu\nu} \,=\,  \mathcal{D}_{\mu}A_{\nu} \,-\, \mathcal{D}_{\nu}A_{\mu} \,+\, ig \left[A_{\mu}, A_{\nu} \right] ~.  \] The equation of motion can be derived easily in the matrix form

\[ \delta\mathcal{W} \,=\, -\frac{1}{2}\int d^4x \, \text{Sp}\left(\mathcal{F}_{\mu\nu}\delta\mathcal{F}^{\mu\nu} \right) ~,  \] where \[ \delta\mathcal{F}^{\mu\nu} \,=\, \mathcal{D}^{\mu}\delta A^{\nu} \,+\,ig\delta A^{\mu}A^{\nu} \,+\, igA^{\mu}\delta A^{\nu} \,-\, \left(\mu \leftrightarrow  \nu \right) ~. \] Because the antisymmetry of $\mathcal{F}^{\mu\nu}$

\begin{eqnarray} \delta\mathcal{W} \,=\,   - \int d^4x \, \text{Sp}\left\{\mathcal{F}_{\mu\nu}\left( \mathcal{D}^{\mu}\delta A^{\nu} \,+\,ig\delta A^{\mu}A^{\nu} \,+\, igA^{\mu}\delta A^{\nu} \right) \right\} ~.  \nonumber \end{eqnarray} The first term in this equation can be integrated by parts, throwing away the surface terms and using the cyclic properties of the {\tt Spur}, it reduces to    

\[ \delta\mathcal{W} \,=\,  \int d^4x \, \text{Sp}\left\{\left( \mathcal{D}^{\mu} \mathcal{F}_{\mu\nu}  \,+\,ig \left[A^{\mu},  \mathcal{F}_{\mu\nu} \right] \right)\delta A^{\nu} \right\} ~,  \] from which we read off the equation of motion in the matrix form 

\[ \mathcal{D}^{\mu} \mathcal{F}_{\mu\nu}  \,+\,ig \left[A^{\mu},  \mathcal{F}_{\mu\nu} \right]  \,=\, 0~.  \] If the coupling of Yang-Mills field to the external source is introduced: $ \int d^4x \, \text{Sp}\left( A^{\nu}J_{\nu}\right)$, then one obtains the equation of motion

\[ \mathcal{D}^{\mu} \mathcal{F}_{\mu\nu}  \,+\,ig \left[A^{\mu},  \mathcal{F}_{\mu\nu} \right]  \,=\, J_{\nu}~.  \] 

Let us focus on $U(1)$ case; the Maxwell equations get modified as

\[ \mathcal{D}^{\mu} \mathcal{F}_{\mu\nu}   \,=\, J_{\nu}~.  \] To see how the Poisson equation gets modified, let us consider a static source: $J_0(\mathbf{x}), J_j=0$. One obtains the equation

\begin{equation}\label{modpuasonisgantoleba} \mathcal{D}^{j}\mathcal{F}_{j0} \,=\, \mathcal{D}^{j}\mathcal{D}_{j}A_0 \,=\, -\frac{\Delta}{\left(1+\beta \Delta\right)^2}\, A_0 \,=\,  J_0(\mathbf{x}) ~. \end{equation} To make a proper analysis of this equation let us recall that the operator $\Delta \left(1+\beta \Delta\right)^{-2}$ arises as a result of using the deformed momentum operator in field theory (see Eq.\eqref{damukhtscalqmedebamod}). Therefore, the Fourier representation of the field includes the cutoff $\mathbf{p}^2 < \beta^{-1}$ (see the section Introduction) 

\[ A_0(\mathbf{x}) \,=\, \frac{1}{\left(2\pi\right)^{3/2}} \int\limits_{\mathbf{p}^2 < \beta^{-1}} d^3p \, e^{-i\mathbf{p}\cdot\mathbf{x}} \mathcal{A}_0(\mathbf{p})~, \] and similar cutoff is implied for the current as well

\[ J_0(\mathbf{x}) \,=\, \frac{1}{\left(2\pi\right)^{3/2}} \int\limits_{\mathbf{p}^2 < \beta^{-1}} d^3p \, e^{-i\mathbf{p}\cdot\mathbf{x}} \mathcal{J}_0(\mathbf{p})~. \]

\noindent  For this reason, one infers that the source $J_0$ cannot be localized beneath the Planck length (see the section Introduction). The solution of Eq.\eqref{modpuasonisgantoleba} reads as  

\[ A_0(\mathbf{x}) \,=\, \frac{1}{\left(2\pi\right)^{3/2}} \int\limits_{\mathbf{p}^2 < \beta^{-1}} d^3p \, e^{-i\mathbf{p}\cdot\mathbf{x}} \, \frac{\left(1 -\beta \mathbf{p}^2 \right)^2\mathcal{J}_0(\mathbf{p})}{\mathbf{p}^2} ~. \]

\noindent For the source represented by the following cutoff version of the $\delta$ function

\[ J_0(\mathbf{x}) \,=\, \frac{1}{\left(2\pi\right)^{3/2}} \int\limits_{\mathbf{p}^2 < \beta^{-1}} d^3p \, e^{-i\mathbf{p}\cdot\mathbf{x}} ~, \] the expression for $A_0$ can be found in \cite{Maziashvili:2011dx}. Let us notice, that the corrected Newtonian potentials due to deformed dispersion relations were studied in a few papers \cite{Tkachuk:2007zz, Helling:2007zv, AmelinoCamelia:2010rm} and recently in \cite{Moayedi:2013nxa, Moayedi:2013nba}.

In order for $A_{\mu}$ to have the equation similar to the Eq.\eqref{modzraobisgantoleba}, one has to require a subsidiary condition $\mathcal{D}^{\mu}A_{\mu}=0$; then for the free field the equation of motion takes the form $\mathcal{D}^{\mu}\mathcal{D}_{\mu}A_{\nu} = 0$. This insures to have similar dispersion relations for scalar and vector particles. But now the condition $\mathcal{D}^{\mu}A_{\mu}=0$ implies two polarization degrees of freedom instead of four.

\section{Classical limit for ml-QM}

\subsection{Minimum-length deformed classical dynamics}

In ml-QM the Hamilton's operator takes the form given by Eq.\eqref{Hamiltonian}. For the velocity operator one finds\footnote{The calculation is easy to do in the $\mathbf{p}$ representation: $\widehat{\mathbf{p}} = \mathbf{p},\, \widehat{\mathbf{r}} = i\partial/\partial \mathbf{p}$.} \cite{LL3}

\begin{eqnarray} \widehat{\mathbf{v}} \,\equiv\, \dot{\widehat{\mathbf{r}}} \,=\, i\left[\widehat{\mathcal{H}} ,\,  \widehat{\mathbf{r}} \right] \,=\, \sum\limits_{n =0}^{\infty} \frac{ (1+n)\beta^n }{2m} \left[  \widehat{\mathbf{p}}^{2(n+1)} ,\,  \widehat{\mathbf{r}} \right]  \,=\,     \frac{ \widehat{\mathbf{p}}}{m}  \sum\limits_{n =0}^{\infty} (1+n)^2\beta^n \widehat{\mathbf{p}}^{2n} \,=\,   \frac{ \widehat{\mathbf{p}}}{m}  \, \frac{1 \,+\,\beta \widehat{\mathbf{p}}^2}{\left( 1 \,-\, \beta \widehat{\mathbf{p}}^2 \right)^3} ~. \nonumber \end{eqnarray}

To pass to the classical mechanics one has to replace the commutator with the Poisson bracket \cite{LL3} 

\begin{equation} \frac{i}{\hbar}\, \left[\widehat{f}_1,\, \widehat{f}_2\right] ~ \rightarrow ~ \left\{f_1,\,f_2 \right\} ~. \nonumber \end{equation} So we arrive at the Hamilton equations 

\[ \dot{\mathbf{r}} \,=\, \left\{\mathcal{H},\, \mathbf{r} \right\} ~,~~~~ \dot{\mathbf{p}} \,=\, \left\{\mathcal{H},\, \mathbf{p} \right\} ~, \] with the standard Poisson brackets \cite{LL1} 

\begin{eqnarray} \left\{x_i,\,x_k \right\} \,=\, 0~, ~~  \left\{p_i,\,p_k \right\} \,=\, 0~, ~~\left\{p_i,\,x_k \right\} \,=\, \delta_{ik}~, \nonumber \end{eqnarray} 

\noindent and modified Hamiltonian, see Eq.\eqref{Hamiltonian},

\begin{eqnarray}\label{klasikuriplqmhamiltoniani} \mathcal{H} \,=\, \frac{\mathbf{p}^2}{2m \left( 1\,-\, \beta \mathbf{p}^2 \right)^2} \,+\, V(\mathbf{r}) ~. \end{eqnarray}

\subsection{Kepler's problem}
\label{Kepler}

In classical dynamics there are only two cases in which all the bounded orbits in central fields are closed,  namely, $V(r) = ar^2, \, a >0$ and $V(r) = -\alpha/r, \, \alpha >0$. The Hamiltonian for describing the motion in a Newtonian potential gets modified as (see Eq.\eqref{klasikuriplqmhamiltoniani}) 

\begin{eqnarray}  \mathcal{H} = \frac{\mathbf{p}^2}{2m \left( 1\,-\, \beta \mathbf{p}^2 \right)^2} - \frac{\alpha}{r} =  \mathcal{H}_0 + \sum\limits_{n =1}^{\infty} \frac{(1+n)\beta^n \mathbf{p}^{2(n+1)}}{2m}  ~. \nonumber \end{eqnarray} The angular momentum $\mathbf{L} = \mathbf{r}\times \mathbf{p}$ is still conserved 

\begin{eqnarray} \frac{d\mathbf{L}}{dt} \,=\, \left\{\mathcal{H},\, \mathbf{L} \right\} \,=\,   \mathbf{r}\times  \left( - \frac{\alpha}{r^2} \, \frac{\mathbf{r}}{r} \right)  \,+\, \frac{d}{dp} \left( \frac{p^2}{2m \left( 1\,-\, \beta p^2 \right)^2} \right) \frac{\mathbf{p}}{p} \times \mathbf{p} \,=\, \mathbf{0}~. \nonumber  \end{eqnarray} So, the motion occurs in the plane perpendicular to $\mathbf{L}$. Let $x,\,y$ be rectangular coordinates in this plane. Then from the equations of motion one gets

\begin{eqnarray}
\dot{x} \,=\, \left\{\mathcal{H},\, x \right\} \,=\, \frac{p_x}{m} \, \frac{1 \,+\, \beta p^2}{\left(1\,-\,\beta p^2 \right)^3}~, ~~~~  \dot{y} \,=\, \left\{\mathcal{H},\, y \right\} \,=\, \frac{p_y}{m} \, \frac{1 \,+\, \beta p^2}{\left(1\,-\,\beta p^2 \right)^3}~, \nonumber
\end{eqnarray} and correspondingly 

\begin{eqnarray}\label{sichkareimpulsikavshiri} m\sqrt{\dot{x}^2 \,+\,\dot{y}^2} \,=\, p\, \frac{1 \,+\, \beta p^2}{\left(1\,-\,\beta p^2 \right)^3} \,=\, p \,+\, 4\beta p^3 \,+\,  9\beta^2 p^5 \,+\, 16\beta^3 p^7 \,+\, 25\beta^4 p^9 \,+\, O\left(\beta^5\right) ~.~ \end{eqnarray} Can we restrict ourselves to the first order in $\beta$ in discussing the motion of solar system planets? The validity of this approximation means that $\beta m^2v^2 \ll 1 $. Taking $\beta \sim l_P^2 \equiv m_P^{-2}$, one observes that since the mean orbital velocities of the planets are by a few orders of magnitude smaller than the light velocity in vacuum but their masses are much more greater as compared to $m_P$, the above condition is not satisfied. Using the mean orbital velocities and masses of planets one gets that the order of magnitude for $\beta m^2v^2 $ varies from $10^{54}$ to $10^{64}$. For the motion of moon around the earth this quantity is of the order of $10^{47}$.

Let us focus on the motion of the moon in earth's gravitational field. The energy expression (which is certainly a conserved quantity) takes the form

\begin{equation}\label{srulienergia}
E   \,=\, \frac{\mathbf{P}^2}{2M_{moon}} \,-\, \frac{m_P^{-2} M_{earth}M_{moon}}{r} ~,
\end{equation} where $M_{earth} =5.9736\times 10^{24}$\,kg and $M_{moon} =0.07349\times 10^{24}$\,kg \cite{NASA}. Using $m_P \approx 2.17651\times 10^{-8}$\,kg one finds $m_P^{-2} M_{earth}M_{moon} \approx 0.2017 \times 10^{40}$. The orbital velocities of the moon at the perigee $r_{+}= 363300$\,km and apogee $r_{-}= 405500$ km are $v^{(0)}_{+}= 1.076$\,km$/$s and $v^{(0)}_{-} = 0.964$ \,km$/$s respectively \cite{NASA} (we introduced the notation $v^{(0)}$ to distinct between standard and modified cases). From Eq.\eqref{sichkareimpulsikavshiri} one observes that for this range of velocities $p^2$ is close to $1/\beta$ with a great accuracy. This fact allows to somewhat simplify the Eq.\eqref{sichkareimpulsikavshiri}

\begin{eqnarray}\label{gamartivebuligamossichimpul}
m v  \,=\, p\, \frac{1 \,+\, \beta p^2}{\left(1\,-\,\beta p^2 \right)^3} \,=\, P^3 \left(\beta \,+\, \frac{1}{p^2} \right) \,\approx\, 2\beta P^3~. \end{eqnarray} Now taking $v_{-}=v^{(0)}_{-}$ and $\beta = m_P^{-2}$, from Eqs.(\ref{srulienergia}, \ref{gamartivebuligamossichimpul}) one finds

\begin{eqnarray}\label{bologantoleba}
\left(\frac{v_{+}}{2} \right)^{2/3} \,=\, \left(\frac{v^{(0)}_{-}}{2} \right)^{2/3} \,+\,  \left(\frac{M_{moon}}{m_P}\right)^{4/3} \left(\left[v^{(0)}_{+}\right]^2 \,-\, \left[v^{(0)}_{-}\right]^2 \right) ~.
\end{eqnarray} Substituting the above cited quantities in Eq.\eqref{bologantoleba} one finds (in natural units) 

\[ v_{+} \simeq 10^{43}~. \]

\section{Discussion}

We address a number of questions concerning the broad class of ml-QM admitting Hilbert space representation in which the deformation is completely ascribed to the momentum operator \cite{Maziashvili:2012dd}. From the very outset it is obvious that $\mathsf{QM}$ underlying the minimum-length deformed uncertainty relation should forbid the localization of wave function in the configuration space beneath the length scale $\sqrt{\beta}$. In a particular Hilbert space representation of ml-QM used throughout this paper, this fact is reflected by the appearance of cutoff on $p$. Otherwise speaking, the Hilbert space of state vectors is restricted to the functions admitting the representation \eqref{chamochrilitsarmodgena}. So, loosely speaking, the singular behaviour of the wave function is uniquely avoided from the very outset, it not a dynamical effect; the corrections to the Hamiltonian do not provide any other mechanism for avoiding the wave-function singularities.

Let us notice that for ensuring $\delta x \gtrsim \sqrt{\beta}$ it suffices to assume the cutoff $p \lesssim \beta^{-1/2}$ and abandon additional deformation of the theory. To make this point clearer, let us start from a first order Lagrangian describing the motion of a single particle   

\begin{equation}\label{firtsorder}
\mathcal{L} \,=\, p_i\dot{x}^i \,-\, \mathcal{H}(\mathbf{x}, \mathbf{p}) ~,~~i=1,2,3.
\end{equation} Upon introducing a new variable for the momentum $\mathbf{p} \rightarrow \mathbf{P}$, the Eq.\eqref{firtsorder} and the corresponding equations of motion take the form 

 \begin{eqnarray}\label{newmomentum}
 \mathcal{L} \,=\, p_i\left(\mathbf{P}\right)\dot{x}^i \,-\, \mathcal{H}\left(\mathbf{x}, \mathbf{p}\left(\mathbf{P}\right)\right) \,=\, p_i\left(\mathbf{P}\right)\dot{x}^i \,-\, \widetilde{\mathcal{H}}\left(\mathbf{x}, \mathbf{P}\right)  ~, \nonumber \\  \frac{\partial p_i}{\partial P_k}\dot{P}_k \,=\,-\, \frac{\partial \widetilde{\mathcal{H}}}{\partial x^i}~,~~~~ \frac{\partial p_i}{\partial P_k}\dot{x}^i \,=\, \frac{\partial \widetilde{\mathcal{H}}}{\partial P_k}~. ~~~~~~~
\end{eqnarray} Here the matrix $\partial p_i/\partial P_k$ is invertible since we assumed that the change of variables is well defined. So the Eq.\eqref{newmomentum} can be written as 

 \begin{eqnarray}\label{inverted}
 \dot{P}_k \,=\,- \, \frac{\partial P_k}{\partial p_i} \frac{\partial \widetilde{\mathcal{H}}}{\partial x^i}~,~~~~ \dot{x}^i \,=\, \frac{\partial P_k}{\partial p_i}\frac{\partial \widetilde{\mathcal{H}}}{\partial P_k}~. \end{eqnarray} One can now put the Eq.\eqref{inverted} in the form 
 
\begin{eqnarray}\label{}
\dot{P}_k \,=\, \left\{\widetilde{\mathcal{H}}, P_k \right\} = \frac{\partial \widetilde{\mathcal{H}}}{\partial x^i} \left\{x^i, P_k \right\} \,+\, \frac{\partial \widetilde{\mathcal{H}}}{\partial P_i} \left\{P_i, P_k \right\} ~, \nonumber \\ \dot{x}^i \,=\, \left\{\widetilde{\mathcal{H}}, x^i \right\} \,=\, \frac{\partial \widetilde{\mathcal{H}}}{\partial x^k} \left\{x^k, x^i \right\} \,+\, \frac{\partial \widetilde{\mathcal{H}}}{\partial P_k} \left\{P_k, x^i \right\} ~, \nonumber \end{eqnarray} under assumption that the brackets are defined as 

\begin{eqnarray}\label{vitsyebtakedan}
 \left\{P_i, P_k \right\} \,=\,0 \,=\,  \left\{x^i, x^k \right\}~,~~   \left\{P_k, x^i \right\} \,=\, \frac{\partial P_k}{\partial p_i} ~. 
\end{eqnarray} As the change of variables $\mathbf{p} \rightarrow \mathbf{P}$ entails the modification $\mathcal{H} \rightarrow \widetilde{\mathcal{H}}$, in doing the reverse procedure (starting from Eq.\eqref{vitsyebtakedan}) one might first make the transformation $\mathbf{P} \rightarrow \mathbf{p}$ and then use the standard Hamiltonian. Following this way, in the case of ml-QM one would arrive at the standard Hamiltonian with a cutoff on $p$.

The following important step is to examine the meaning of the wave-function and some related issues. First we ask how to introduce electromagnetic interaction in ml-QM. To answer this question, we take into account that the standard rule for inclusion of electromagnetism in quantum mechanics         $\nabla \rightarrow \nabla + iq\mathbf{A}$ comes from the fact that classically the coupling of the electromagnetism to the particle leads to the replacement $\mathbf{P}\rightarrow \mathbf{P} + q\mathbf{A}$; see Eqs.(\ref{ararelnatselmagveli}, \ref{ararelnatsurtgareveltan}). Taking the same classical picture as a starting point, one notices that in the case of ml-QM the transition to quantum formalism is achieved with the transcription $\mathbf{P} \rightarrow -i\nabla/\left( 1+\beta\Delta \right)$ and therefore an interaction with an external electromagnetic field is introduced by the substitution $\nabla/\left( 1+\beta\Delta \right) \rightarrow \nabla/\left( 1+\beta\Delta \right) + iq\mathbf{A}$; see Eqs.(\ref{deformedopmultid}, \ref{Hamiltonianem}). That is, we uniquely follow the way: $\widehat{\mathbf{P}} \rightarrow \widehat{\mathbf{P}} + q\mathbf{A}$. Generalization of this rule to the case of four-potential $(A^0, \mathbf{A})$ is straightforward as the operator $\widehat{P}^0$ is unmodified; see Eqs.(\ref{damukhtscalqmedebamod}, \ref{scactionone}, \ref{elmagvelischartvistsesi}). Let us notice that our approach differs from that one suggested in \cite{Hossenfelder:2003jz, Kober:2010sj} to use the rule                                 $\partial_{\mu} \rightarrow \partial_{\mu} + iqA_{\mu}$ for all $\partial_{\mu}$ operators in the deformed momentum. The proposal of \cite{Hossenfelder:2003jz, Kober:2010sj} leads to the infinite number of interaction terms, that is, the interaction term is represented as a series of powers of $\beta$. In our case it is easy to see that one arrives at a finite number of interaction terms, see Eq.\eqref{damukhtuliskalari}.

Next, we ask how to derive the expression of electric current in ml-QM. Following the Landau-Lifshitz derivation, see Eq.\eqref{LLgamosakhulebakmdenistvis}, the electric current is defined by means of the background electromagnetic field $\mathbf{A}$ as   

\[ \frac{\delta \langle H \rangle_{\Psi}}{\delta \mathbf{A}(\mathbf{x})} \,=\,   \mathbf{J}(\mathbf{x}) ~. \] But the current derived this way does not satisfy the continuity equation and therefore the standard probabilistic interpretation of the wave-function becomes obscure, Eq.\eqref{kvantmechdeni}. On the other hand, one can define the conserved current immediately from the Schr\"odinger equation, but then its physical interpretation becomes obscure, Eqs.(\ref{shenakhvadidenischrgantolebidan}, \ref{shenakhvadikvanmechdeni}). Similarly, in the case of ml-QFT Noether's current, see Eqs.(\ref{denisnulovanikomponenta}, \ref{denissivrtsulikomponenta}), does not coincide with the one that comes immediately from the action functional after inclusion of the electromagnetic field, Eq.\eqref{damukhtuliskalari}.

Another interesting question is the gauge invariance; the above discussion manifests the violation of gauge invariance in ml-QM, ml-QFT (last paragraph in section II and the Eq.\eqref{damukhtuliskalari}). Therefore, the degrees of freedom that are reduced due to gauge invariance in the standard case are now physical. But, nevertheless, in order to have the same dispersion relation for all particles, one has to impose on the electromagnetic field the additional condition (last paragraph in section III).

Finally (section IV) we point out dangerous implications of ml-QM for classical physics. Namely, its classical counterpart manifests huge effects when applied for the motion of planets. Similar observations were made in \cite{Benczik:2002tt} in the framework of different Hilbert space representation of ml-QM and also in \cite{Silagadze:2009vu}. The latter paper used the above discussed representation but consideration was restricted to the first order in $\beta$ that for our purposes can not be considered a good approximation.

\acknowledgments  Useful discussions with Zurab~K.~Silagadze are kindly acknowledged. M.~M. is indebted to Marcus Bleicher and Piero Nicolini for the hospitality at the {\it Frankfurt Institute for Advanced Studies} where the paper was extensively modified by adding a substantial amount of new material. This research was supported in part by the Shota Rustaveli National Science Foundation under contract number 31/89, the {\tt DAAD} research fellowship for university teachers and researchers and the grant: I/84600 from {\tt Volkswagen Stiftung}.


\end{document}